# Luminescence from ZAIS quantum dots: radiative or nonradiative

*Yasaman Moradi[1], Rene Zeto[1], Andrea M. Armani[1,*]*

[1] Mork Family Department of Chemical Engineering and Materials Science, University of Southern California, Los Angeles, California 90089, USA

**Corresponding Author**

*Andrea M. Armani, Email: armani@usc.edu

Quantum dots have found applications across the semiconductor and biotechnology industries, improving energy efficiency, color purity, and imaging capability. However, the standard quantum dot relies on a heavy metal architecture, which has undesirable environmental impacts. Thus, alternative material systems are desirable. One strategy is the ZnS-AgInS$_2$ (ZAIS) quantum dot which has a metal dichalcogenide chemical structure. Unlike heavy metal quantum dots, these nanoparticles luminesce even when defects are present, and their emission wavelength is tuned by varying the relative composition. Past works have linked the source of this luminescence to a defect-assisted transition recombination process, but the precise mechanism is still unclear. In this work, we investigate the physics of this defect-assisted transition by systematically varying the concentration of defects through a simple quick-cool thermal annealing process which provides control over the crystalline disorder of the nanoparticle. Using a combination of ultraviolet photoelectron spectroscopy and fluorescence measurements, information about the nature of the defect-assisted transition is obtained. In contrast to previous work, we find that the ZAIS quantum dot luminescence has an electronic-based rather than defect-assisted transition mechanism.

**TOC GRAPHICS**

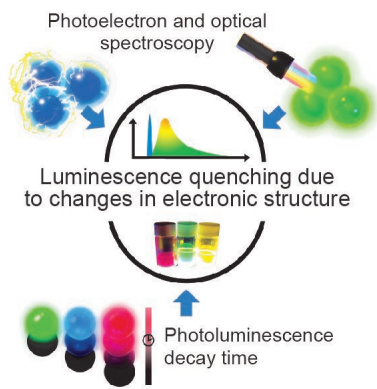

**KEYWORDS**. Quantum dot, Photoluminescence, Luminescence, metal dichalcogenide

Over the last several decades, quantum dots have evolved from being a tool to study quantum phenomena to serving as a platform technology for bioimaging and displays [1–3]. This shift from basic science to applications is directly tied to their photostability [4], quantum yield [5], and photon lifetime [6], and all of these metrics are connected to the material system [7].

Due to its ease of synthesis, high performance, and extensive characterization data, one commonly used chemical composition is cadmium chalcogenide with a protective zinc chalcogenide overcoating[8,9]. Since its initial conception [10], researchers have improved several metrics including color quality and photon lifetime, and the emission wavelength is controlled by the quantum confinement effect which dictates that the wavelength is governed by the particle size [11]. However, as a toxic heavy metal, Cadmium has negative impact on the environment [12].

To overcome these toxicity issues, a new family of quantum dot materials, known as quaternary heavy-metal-free (QHMF) quantum dots, have been developed [13]. These nanoparticles are based on a metal dichalcogenide, $A_xB_yS_2$ (A = Ag or Cu, B = In or Ga) structure [14,15]. In contrast to the previous crystalline binary and ternary structures, metal dichalcogenides can tolerate a wide range of chemical compositions or stoichiometries; that is, the crystal lattice can still form and retain useful optical properties even with a severe deficiency or surplus of each constituent element [16]. Although the size effect is still present in QHMF quantum dots, the emission wavelength is primarily determined by compositional factors, which can be more precisely controlled than size. Additionally, in QHMF quantum dots, there are four constituent elements [17]. Therefore, the tunability is not only more precise but there are also more degrees of freedom to the system, providing multiple paths for material optimization.

Among the different QHMF systems, ZnS-AgInS$_2$ quantum dots, or ZAIS, have demonstrated a range of useful applications from solar cells to fluorescent probes [18,19]. To optimize the performance of these quantum dots, the effects of changing the Ag:In ratio and of changing the ZnS shell thickness on the luminescence intensity of ZAIS QDs have been investigated in separate studies [20,21]. The results demonstrate that the quantum yield does not linearly correlate with Ag:In ratio, and there is an optimum ratio that can give the strongest luminescence signal. Additionally, increasing the ZnS shell thickness by adjusting the molar ratio of Zn to $AgInS_2$, from 2:1 to 4:1 has been shown to further improve the fluorescent intensity of the quantum dots.

To understand this behavior from a mechanistic perspective, past works have investigated the role of carrier recombination at defect sites. According to these studies, defect sites provide near band edge traps for carriers that can later recombine to relax and emit a photon. Thus, the ZAIS luminescence is primarily a defect-assisted process [22]. To experimentally study this behavior, previous work relied on a slow thermal treatment followed by a recrystallization process which allowed the ions in the lattice to rearrange to thermodynamically favorable states. It was hypothesized that this process reduced the defect concentration and thus the likelihood that carriers radiatively recombine [22,23]. In agreement with the proposed mechanism, the predicted emission quenching was observed [23]. However, the complementary material system in which defect states are intentionally locked-in has not been investigated. Thus, the proposed mechanism has yet to be definitively established.

In this work, we synthesize ZAIS quantum dots using a quick-cool annealing method. This technique does not support the recrystallization process, allowing the defect-assisted luminescence hypothesis to be systematically studied using a combination of spectroscopic techniques. Optical absorption spectroscopy and fluorometry were used to measure the defect concentration and the quantum yield. Surprisingly, even with the intentionally poor crystallization, the same luminescence quenching was observed. These results directly contradict the previously hypothesized defect-assisted luminescence mechanism. To further understand the nature of the luminescence quenching, ultraviolet photoelectron spectroscopy and time-correlated photoluminescence decay measurements were performed. When the measurement results are examined holistically, it became apparent that a shift in the electronic structure near the bandgap has occurred, providing strong evidence that this thermally induced quenching causes a fundamental change in the electronic structure of the material.

The ZAIS quantum dots were synthesized according to the hot injection strategy (Fig. 1a) [24]. After purification, ZAIS quantum dot films were deposited on glass slides. The samples were subsequently annealed using the quick-cooled method. To study the dependence on annealing temperature, the temperature was varied from 25 °C to 240 °C. Additional details are contained in the supplemental information (SI).

To determine the chemical composition and confirm the optical activity, the synthesized ZAIS quantum dots prior to quick-cool annealing were characterized using X-ray photoelectron spectroscopy (XPS), optical absorption spectroscopy, and fluorescence spectroscopy. Based on the XPS data, the chemical composition of the ZAIS quantum dots is 3.6% Ag, 14.5% In, 23.6% S, and 58.2% Zn (Fig. S1). This composition is held constant throughout all measurements. The optical absorption and emission behavior are similar to prior ZAIS quantum dots (Fig 1 b) [24]. A qualitative analysis of a series of thin film samples after undergoing the quick cool annealing process reveals a rapid fluorescence intensity decrease at higher anneal temperatures (Fig. 1c). This finding also mirrors the results from prior work [23].

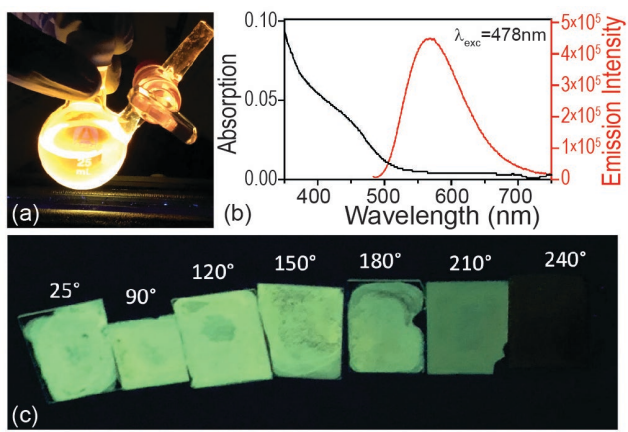

**Figure 1.** (a) Image of ZAIS quantum dots immediately upon completion of reaction. The excitation source is a UV lamp (365 nm). (b) ZAIS quantum dot absorption (black) and emission (red) in chloroform pre-anneal. (c) The emission is clearly visible from the ZAIS quantum dot thin films deposited on glass slides However, as the quick-cool anneal temperature increases, the emission decreases. The excitation source is a UV lamp.

To quantitatively correlate the optical behavior with the concentration of defect states in the ZAIS quantum dots, optical absorption spectroscopy measurements are performed on the thin film samples to measure the band gap energy ($E_g$). Typically, semiconductor quantum dots exhibit a sharp edge in the absorption spectrum at $E_g$. However, the presence of defects results in a significant softening of this edge. By measuring the absorption spectra of the particles and looking at the sharpness of that transition, the amount of disorder can be quantified as an Urbach energy ($E_u$) [25]. Details of this calculation are included in the SI. As shown in Fig. 2, the results from the quick-cooled annealing treatment demonstrates that the annealing process successfully changed the Urbach energies of the samples. Notably, higher quick-cool annealing temperatures result in over a 10x increase in $E_u$ values, indicating an increase in disorder, as would be expected by the quick-cool synthesis process.

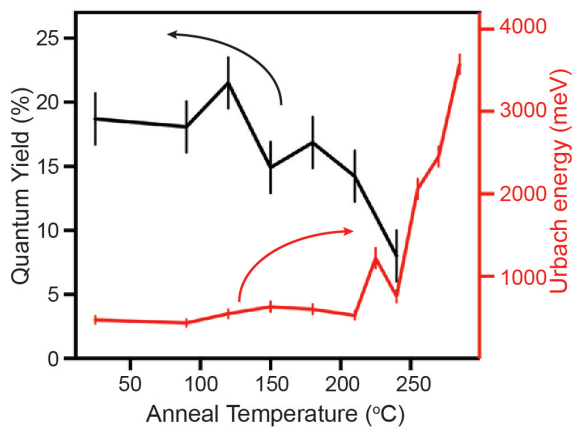

**Figure 2**. Correlating the formation of disordered lattice which is represented as Urbach energy with observed variations in ZAIS quantum yield at different annealing temperatures

The previously established mechanism postulated that the defect concentration, which is related to the $E_u$, should be independent of the fluorescent quantum yield. Given the clear visual evidence of a decrease in fluorescence (Fig 1c), the quantum yield was quantitatively analyzed, and the results revealed that the quantum yield values approach zero at higher temperatures. Notably, the signals from samples annealed above 240 °C were not detectable. In other words, the photoluminescence was completely quenched when the crystalline lattice was more disordered, as parameterized by the samples' Urbach energies. This finding directly contradicts the prior hypothesis. Given that the chemical composition was constant across all samples studied, one possible explanation for these results is that the quick-cool method effects the electronic structure of the ZAIS quantum dots.

To investigate this new hypothesis and probe the electronic structure, ultraviolet photoelectron spectroscopy (UPS) data was obtained for the annealed samples (Fig. 3). UPS is a fine-resolution technique for measuring the binding energies of outer shell (e.g., valence) electrons on the surface of a material. As such, UPS measurements analyze the density of states at band-gap energies, which are normally too narrow of a band to measure using x-ray photons in traditional XPS measurements [26–28]. As can be seen in Fig. 3, the UPS results show that lower temperature samples present an exponentially decaying band tail, which is consistent with the optical Urbach energies measured. However, a significant change occurs as the temperature is increased above 200 °C, with a sharper band edge appearing around 2 eV binding energy. In particular, the shoulder that appears in the spectrum of the samples annealed higher than 200 °C seems to suggest the consolidation of states in a band, rather than the poorly defined edge expected in amorphous semiconductors. The results in Fig 3 along with the additional data in Fig. S2 provide evidence that an intrinsic change in the electronic structure of the quantum dots occurs when the samples are annealed above 200 °C. Notably, this temperature point aligns with the temperature at which the photoluminescence decreases and the Urbach energy increases.

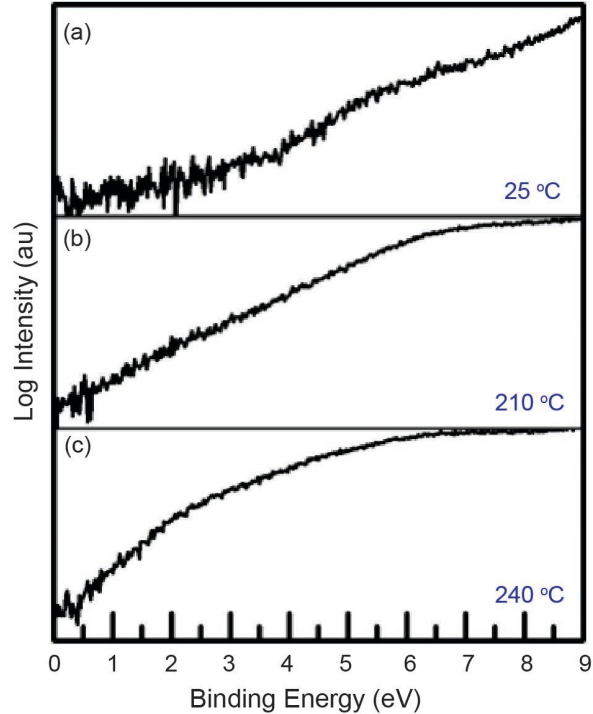

**Figure 3.** Ultraviolet photoelectron spectroscopy (UPS) data of the quick-cooled annealed samples, indicating a significant electronic structural change at higher temperatures.

To bring insight into the nature of the excited states, the photoluminescence lifetimes of all samples prepared were measured post-anneal using time-resolved photoluminescence spectroscopy (Fig. S3). A two-term exponential model provided the most accurate fit for the recorded decay curves, indicating that two excited states govern the emission process. Fig. 4 shows the variations in one of the excited state lifetimes at different annealing temperatures. According to this data, at annealing temperatures less than 200 °C, the lifetime of this excited state does not significantly fluctuate which supports the previously hypothesized donor-acceptor recombination process model [29,30]. However, at higher temperatures, a drastically longer excited state lifetime of 600 ns was observed. This noticeable increase suggests that this excited state is significantly perturbed by the annealing process, and the responsible defect state is affected by thermal processes (e.g., vacancy or interstitial state).

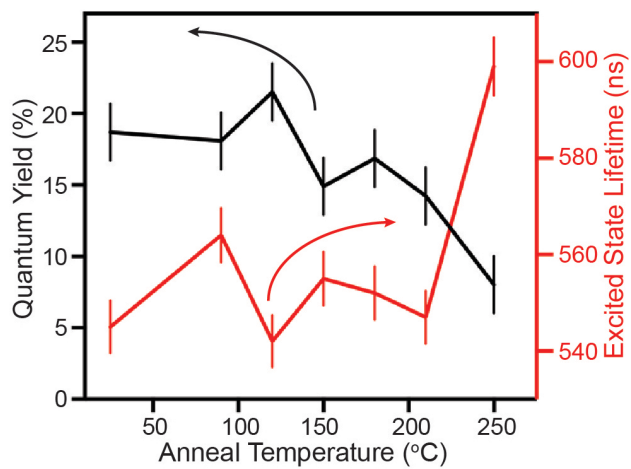

**Figure 4.** Quenched quantum yield with increasing temperatures also correlates with a sharp increase in the excited state lifetime for one of the excited states that is responsible for luminescence.

While the photoluminescence lifetime and quantum yield data both depend on the thermal annealing treatment, the trends are inverse of each other (Fig. 4). This behavior provides additional evidence for the existence of a fundamental electronic structure change. Specifically, the increase in excited state lifetime indicates that another less radiative recombination pathway could have been created during the high temperature quick-cool process. This new pathway could facilitate the decrease in quantum yield despite the high degree of crystalline disorder.

In summary, this study synthesized and characterized ZAIS quantum dots to probe the origin of their photoluminescence experimentally. A quick-cooled thermal annealing treatment was used to manipulate the degree of lattice disorder present in the samples. To parameterize this order, the Urbach energy was measured by recording the tail of the absorption spectrum band edge, and it was found that higher temperature-treated samples had lower degrees of crystallinity due to the quick-cool process used. However, photoluminescence quenching was still observed, indicating that the origin of the defect must be thermal in nature. To confirm that this quenching effect was due to changes in the electronic structure, ultraviolet photoelectron spectroscopy measurements were performed, and they revealed a band shoulder for samples treated at higher temperatures. Lastly, time-resolved photoluminescence decay measurements suggest the possibility of an alternate nonradiative recombination pathway for samples treated at higher temperatures. These findings set the stage for future research investigating compositional dependencies and engineering the optical behavior based on annealing strategies.

ASSOCIATED CONTENT

**Supporting Information**. Additional synthesis and experimental details, Urbach energy calculation details, X-ray photoelectron spectroscopy (XPS) results, Ultraviolet photoelectron spectroscopy (UPS) data, Photoluminescence decay measurements.

AUTHOR INFORMATION

Corresponding Author

Andrea Armani- Mork Family Department of Chemical Engineering and Materials Science, University of Southern California, Los Angeles, California 90089, USA

Authors

Yasaman Moradi- Mork Family Department of Chemical Engineering and Materials Science, University of Southern California, Los Angeles, California 90089, USA

Rene Zeto- Mork Family Department of Chemical Engineering and Materials Science, University of Southern California, Los Angeles, California 90089, USA

**Notes**

The authors declare no competing financial interests.

ACKNOWLEDGMENT

The authors acknowledge the Office of Naval Research for financial support.

*Supplementary Information for:*

# Luminescence from ZAIS quantum dots: radiative or nonradiative

*Yasaman Moradi, Rene Zeto, Andrea M. Armani*

## Contents

## 1　Materials and Methods

### 1.1　Materials

The precursors and solvents along with their manufacturers and associated purities are as follows. All materials were used as arrived with no additional purification.

Materials: Indium (III) acetylacetonate (In(acac)$_3$, 99.99%, Aldrich), Silver nitrate (AgNO$_3$, Ward's science), Sulfur (S, 99.98%, Aldrich), Zinc stearate (10 to 12% Zn basis, Aldrich), 1-octadecene (ODE, 90%, Aldrich), 1-dodecanethiol (DDT, 98%, Aldrich), Oleylamine (OLA, 70%, Aldrich), Oleic acid (OA, 99%, Aldrich), and Trioctylphosphine (TOP, 90%, Aldrich)

### 1.2　Synthesis of ZAIS quantum dots

The hot injection synthesis method involves first creating the AIS core then growing the ZnS shell.

To form the core, 0.1 mmol of AgNO$_3$, 0.4 mmol of In(acac)$_3$, 1.5 mmol of OA, and 25 mmol of ODE were added to a three-neck flask. After purging the flask with N$_2$, the compounds were left stirring for 20 min at room temperature. After 20 min, the temperature was increased to 90°C, and 4 mmol of DDT (purged with N$_2$ gas) was injected into the three-neck flask. The reaction was allowed to proceed for an additional 30 min. The reaction temperature was then increased to 120°C, and a solution of 0.65 mmol of S dissolved in 4 mmol of OLA, which was prepared and stored under N$_2$ in advance, was injected into the three-neck flask and reacted for 3 min.

To form the ZnS shell, a solution of 0.4 mmol of Zn stearate and 0.4 mmol of S dissolved in 0.4 mmol of TOP was mixed in a round bottom flask under N$_2$ purge at RT. This solution was quickly transferred to the three-neck flask through a needle. The temperature was brought up to 180°C and left to react for 2 hours.

To purify the ZAIS quantum dots from the crude oil, 500ml of the unpurified ZAIS and 500ml of Ethanol was added to an Eppendorf centrifuge tube. The particles were well dispersed in the solution by sonicating using an ultrasonic bath. The dispersed solution was centrifuged at 21,000 × g for 5 minutes. The pellet formed after centrifugation was re-dispersed in 1ml of Ethanol. The purification procedure was repeated 3 times. The final ZAIS quantum dot sample was dispersed and stored in Chloroform. Samples were characterized using optical spectroscopy, fluorimetry, and XPS as prepared.

1.3 <u>Preparation of ZAIS thin samples</u>

Glass coverslips are cleaned by quickly rinsing in acetone, methanol, and isopropanol, and then drying on a hotplate (60ºC, ambient environment). To deposit the film, 5 µl of ZAIS stock solution in chloroform was pipetted onto the clean glass slides still on the hotplate, and the chloroform was allowed to evaporate. This process was repeated until a desired coating opacity was achieved. Typically, 750 µl was enough to produce optically dense samples that had good signal to noise for measurements. These dried films on glass coverslips could be used for any optical measurements, such as UV Vis and fluorometry.

The quick-cool annealing process involved a temperature ramp to a variable set point at a fixed rate (5 ºC per 10 minutes), followed by a temperature hold at the variable set point temperature (no treatment, 25, 90, 120, 150, 180, 210, 225, 240, 255, 270, 285 ºC) for each sample. When that anneal process was finished, the samples were quickly removed from the furnace and allowed to quickly cool in air, with the goal of locking in the excited thermal state of the material. Samples were characterized using optical spectroscopy, fluorimetry, UPS, and photon lifetime measurements as prepared.

## 2 Chemical Composition

The chemical composition of the samples was analyzed using a Kratos Axis Ultra DLD X-ray photoelectron spectrometer (XPS). Figure S1 shows measurements of seven different samples from the synthesis batch used in results presented here to demonstrate consistency and uniformity. Based on this data, the chemical composition of the ZAIS quantum dots is 3.6% Ag, 14.5% In, 23.6% S, and 58.2% Zn.

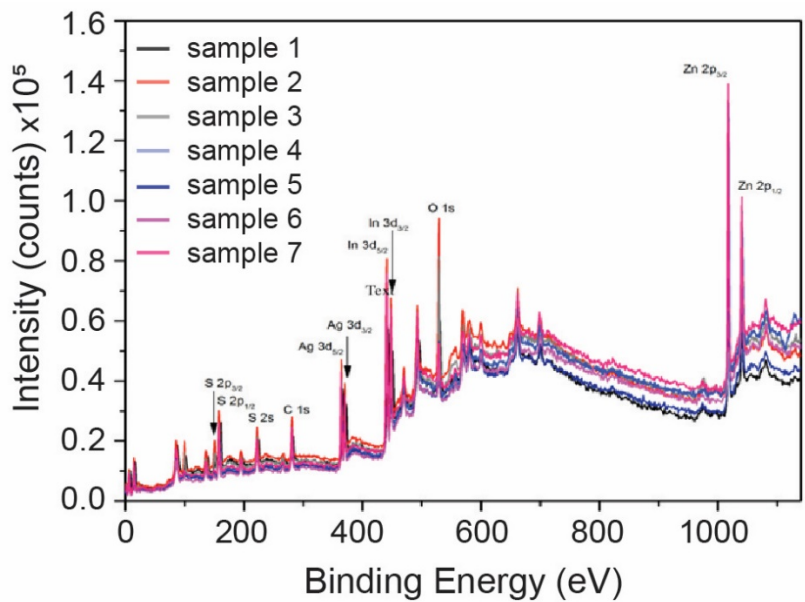

**Fig S1**. X-ray photoelectron spectroscopy (XPS) measurements of seven samples from the same batch representing the chemical and electronic states of synthesized ZAIS quantum dots.

## 3 Absorption and emission measurements

The optical absorption spectra of the samples were measured from 350 nm to 1100 nm in 1 nm increments on a Beckman UV Vis spectrometer. Using this data, the Urbach energies were calculated as detailed in the next section.

Photoluminescence excitation and emission scans were performed using a Horiba FluoroMax-4 spectrofluorometer. Quantum yield measurements were done using an integrating sphere on the same fluorimeter.

## 4 Urbach energy ($E_u$) measurements

Absorption measurements for semiconductor quantum dots typically have a flat region with little to no absorbance for photon energies less than the band gap energy ($E < E_g$), and then a monotonically increasing region for photon energies greater than the band gap energy ($E > E_g$). Tauc coordinates are a useful way to model the absorbance in these difference regimes. When $E > E_g$:

$$\alpha(\nu)h\nu = \alpha_0 \sqrt{h\nu - E_g} \qquad (1)$$

where $\alpha(\nu)$ is the absorption of the particles, $h\nu$ is the incident photon energy, and $\alpha_0$ is a proportionality constant. This expression allows for determination of the band gap from absorption data. When the semiconductor is highly disordered, as in the case of ZAIS quantum dots, the long Urbach tail can be modeled by eq. (2). For $E < E_g$:

$$\alpha(\nu) = \alpha_0 \exp\left(\frac{h\nu}{E_u}\right) \qquad (2)$$

By plotting $\ln\alpha(\nu)$ vs $h\nu$ for ZAIS samples annealed at different temperatures and fitting the slope of the linear region, the $E_u$ was determined[2].

## 5 Ultraviolet Photoelectron Spectroscopy (UPS)

UPS measurements were obtained for each quick cooled sample using a Kratos Axis Ultra DLD photoelectron spectrometer equipped with a Deuterium lamp for UPS measurements (UPS). A subset of the results are in the main text (25 °C, 210 °C, and 240 °C) as Figure 3. All results are included in Figure S2.

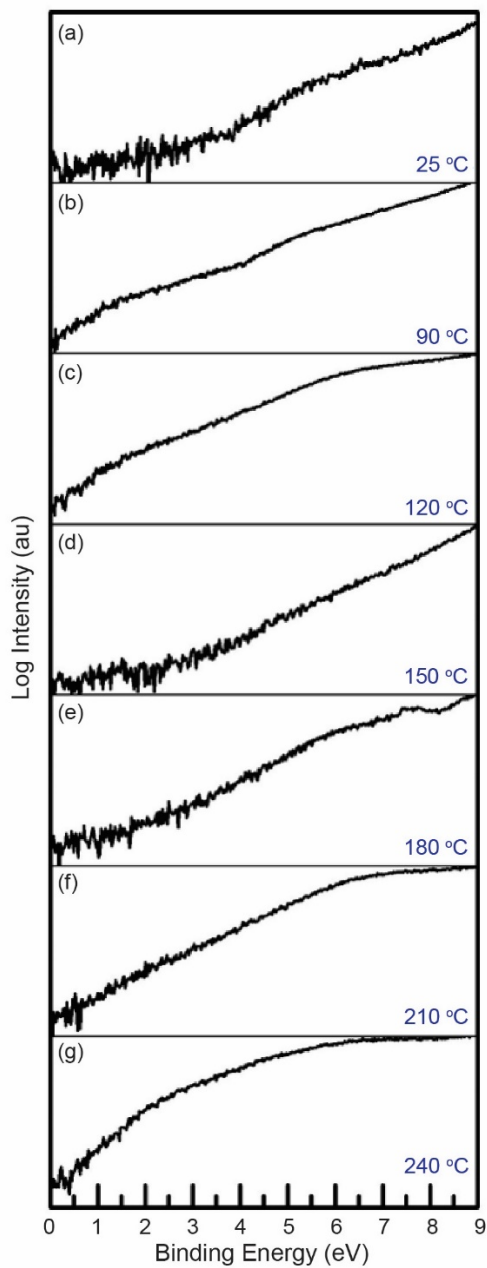

**Fig S2.** Ultraviolet photoelectron spectroscopy (UPS) data of the quick-cooled annealed samples at different annealing temperatures

## 6 Photon lifetime measurements

Time-resolved photoluminescence curves were measured using time-correlated single-photon counting with an IBH Fluorocube instrument equipped with an LED excitation source. The results are presented in Figure S. A two-term exponential model provided the most accurate fit for the recorded decay curves, indicating that two excited states govern the emission process. The values are included in Table S1, and the second lifetime is plotted in Figure 4 in the main text.

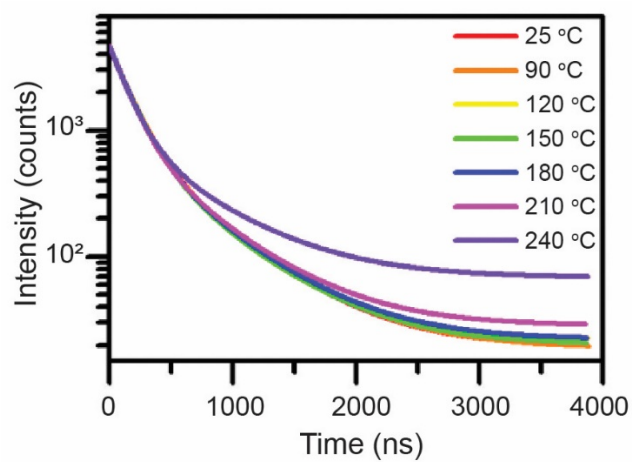

**Fig S3**. Photoluminescence decay measurements of the samples post-anneal using time-resolved photoluminescence.

**Table S1.** Table of all photon lifetime values determined by fitting the data in Figure S3.

| Temperature | Lifetime 1 (ns) | Lifetime 2 (ns) |
|---|---|---|
| 25 °C | 161 | 545 |
| 90 °C | 165 | 564 |
| 120 °C | 158 | 542 |
| 150 °C | 162 | 555 |
| 180 °C | 156 | 552 |
| 210 °C | 151 | 547 |
| 240 °C | 146 | 599 |